\documentclass[aps,preprint,showpacs]{revtex4}

\usepackage{graphicx}

\begin{document}
\title{Energy flow of moving dissipative topological solitons
}
\author{A.V. Gorbach}
\author{S. Denisov}
\author{S. Flach}
\affiliation{Max-Planck-Institut f\"ur Physik komplexer Systeme, 
N\"othnitzer Str. 38, 01187 Dresden, Germany
}


\begin{abstract}
We study the energy flow due to the motion of topological
solitons in nonlinear extended systems in the presence
of damping and driving.
The total field momentum contribution to the energy flux, which reduces
the soliton motion to that of a point particle, is insufficient.
We identify an additional \emph{exchange} energy flux channel mediated
by the spatial and temporal inhomogeneity of the system state.
In the well-known case of a DC external force the corresponding
exchange current is shown to be small but non-zero.
For the case of AC driving forces, which lead to a soliton ratchet,
the exchange energy flux mediates the complete energy flow of the system.
We also consider the case of combination of AC and DC external forces, 
as well as spatial discretization effects.
\end{abstract}

\maketitle

\section{Introduction}

Solitary waves or solitons represent one of the striking and famous
aspects of nonlinear phenomena in spatially extended systems. 
They appear as specific types of localized
solutions of various nonlinear 
partial differential equations and possess several important
properties \cite{Remoissenet}. These properties
include an exponential
localization of energy remaining unchanged during soliton propagation, 
and an elastic scattering of solitons. The latter one, 
i.e. the fact that two solitary waves maintain their identity 
after collisions, inspired the introduction of the term \emph{soliton} \cite{Zabusky}, 
which emphasizes the analogy between these objects and particles. 
Later it has been realized, however, that an essential condition for the particle-like
behavior of solitons is the integrability of the corresponding models. 
But 
the effect of dynamical localization of energy was found to be much more
general, and the
notion of soliton has been extended nowadays to non-integrable models as well. An analogy between 
such "generalized" solitons and particles is less obvious. Not only the scattering process is generally different 
from the elastic scattering of particles, but several other important 
physical effects may appear for solitary waves in non-integrable models. 
An important feature essentially breaking the particle approach to soliton dynamics is connected
with the possibility of excitation of the so-called \emph{shape} \cite{shape_m} or \emph{internal} 
\cite{internal_m} modes: in contrast to a rigid body, 
a soliton may perform shape oscillations while propagating. The mechanism of possible energy exchange between 
the soliton kinetic energy, associated with its translational motion, and the internal energy of shape modes, 
can drastically change the soliton dynamics and leads in some cases to resonances during 
soliton-soliton interactions \cite{shape_m}.
 
Another important aspect of soliton dynamics in non-integrable systems, which we address in this paper, 
concerns the problem of energy transport associated with a directed soliton propagation. Since a soliton
carries some (finite) non-zero energy, it is naturally to assume that its translational motion would result
in a non-zero energy current in the system. One can introduce a \emph{total field momentum}, in analogy to a
particle mechanical momentum, which precisely corresponds to the energy current when considering 
solitons in integrable systems \cite{Remoissenet}.
However, an attempt to extend this momentum-based approach to dissipative solitons may lead
to confusing and misleading results. As an example, we mention here the so-called soliton ratchets: 
the effect of a unidirectional motion of a topological soliton (kink) under the
influence of a spatially homogeneous
and time-periodic external bias with zero mean
\cite{Marchesoni, Salerno, Costantini, Flach, Salerno_Zol, Willis, Molina, Ustinov, Goldobin}. 
As an extension of single particle ratchets \cite{Hanngi}, at a first glance soliton 
ratchets demonstrate at least qualitatively a nice analogy between solitons and particles. 
Indeed, applying the same symmetry analysis 
as for particle ratchets \cite{Flach-sym}, one observes
a directed motion of kinks not only in numerical simulations, 
but also in experiments with annular Josephson junctions \cite{Ustinov, Goldobin}.
However, a surprising and puzzling result appears if one estimates the energy current 
for the soliton ratchet case: simple analytical calculations tell that the 
averaged value of the total field momentum is
strictly equal to zero \cite{we, Quintero}. An analogous effect of non-topological soliton
motion with zero total field momentum
has been reported also for a nonlinear Schr\"odinger equation with parametric driving
in the presence of dissipation \cite{Barashenkov}.
These results show, that  
the total field momentum is no longer an adequate quantity to describe the energy flow
due to soliton dynamics in non-integrable damped and driven
systems.

In this paper we study the energy flow due to soliton dynamics in the
framework of the sin-Gordon model with a spatially homogeneous external
driving and damping. 
We show, that in general the energy current associated with a translational
motion of the topological soliton (kink) consists of two contributions. The
abovementioned total field momentum constitutes only one component of the
total current - the \emph{internal} current. Another path for energy transport
in the system is mediated by an inhomogeneous in time and space energy exchange
between the soliton and the external degrees of freedom (driving force and
damping). The corresponding \emph{exchange} current has no analogy within the
particle description of a moving soliton. Its existence is solely due to 
the spatial extent of the soliton. Even for the case of a time-independent
driving, when shape modes are not excited, the exchange current is small but nonzero.
It is drastically enhanced by the excitation of shape modes via 
a time-periodic driving or a spatial discretization of the system.

The paper is organized as follows. In Sec.~\ref{sec_model} we introduce the driven and damped sin-Gordon model 
and the basic definitions of the energy balance equation and the total field momentum (internal energy current).
In Sec.~\ref{sec_const} the case of soliton motion under the influence of a DC external force is discussed. We
demonstrate, that the internal energy current does not account for the
total energy flow in the system.
We introduce the notion of the exchange current, which
completes the energy flow balance.
A generalization of our approach to the case of soliton motion induced 
by AC driving forces (soliton ratchet) is given 
in Sec.~\ref{sec_rat}. The effects of discretization are discussed in Sec.\ref{sec_discr}. 
Finally, Sec.~\ref{sec_fin} concludes the paper.

\section{The model}
\label{sec_model}

We consider a topological soliton (kink) motion in the 
driven and damped sin-Gordon model \cite{Dodd}, widely used
in the field of soliton ratchets 
\cite{Marchesoni, Flach, Salerno_Zol, Willis, Molina, Ustinov}, which 
describes the dynamics of the superconducting phase difference
across the annular Josephson junction \cite{Ustinov}:
\begin{eqnarray}
\varphi_{tt}-\varphi_{xx}=-\alpha\varphi_{t}-
\sin \varphi+E(t)\;.
\label{sgE}
\end{eqnarray}
This equation describes the evolution of a scalar field $\varphi$ in space and time.
Here and in what follows subscripts $x$ and $t$ denote partial derivatives 
with respect to the corresponding variables.
The parameter $\alpha$ regulates the strength of damping in the system,
and $E(t)$ is an external driving force.
In addition, 
we impose the kink-bearing periodic
boundary condition:
\begin{eqnarray}
\varphi(x+L,t)=\varphi(x,t)+Q\;,\;\varphi_t(x+L,t)=\varphi_t(x,t)\;,
\label{PBC}
\end{eqnarray}
where $Q=2\pi$ is the topological charge and $L$ is the system size. 

The field energy density is given by
\begin{eqnarray}
\rho[\varphi(x,t)] \equiv \rho(x,t)=\frac{1}{2}(\varphi_{t}^{2} +
\varphi_{x}^{2})+1-\cos (\varphi)\;.
\label{density}
\end{eqnarray}
Its dynamics is governed by the following energy balance equation:
\begin{eqnarray}
\rho_{t}=
-j^{I}_{x}-\alpha
\varphi_{t}^{2}+E(t)\varphi_{t}\;.
\label{en_bal}
\end{eqnarray}
Here we introduce the \emph{internal energy current} $J^{I}(t)$ and its density $j^{I}(x,t)$:
\begin{eqnarray}
J^{I}(t)=\int_{0}^{L} j^{I} dx \;,\;
j^{I}(x,t) = - \varphi_{x}
\varphi_{t}\;.
\label{JINT}
\end{eqnarray}
$J^I$ is also known in the literature 
as the total field momentum \cite{Dodd, Joergensen, Olsen, Bonilla}. By differentiating Eq.~(\ref{JINT}) and using
Eq.~(\ref{sgE}) together with the boundary condition~(\ref{PBC})
it is straightforward to show
\cite{Joergensen, Olsen, Quintero} that the internal current $J^I$ satisfies the following ordinary differential equation:
\begin{eqnarray}
J^{I}_{t}(t)= -\alpha J^{I}(t)-Q E(t)\;.
\label{JODE}
\end{eqnarray}

In the absence of an external force $E(t)\equiv 0$ and dissipation $\alpha \equiv 0$ 
(i.e. in the \emph{non-perturbed} integrable sin-Gordon model), Eq.~(\ref{sgE}) with 
the above boundary conditions~(\ref{PBC})
supports the well-known kink solution \cite{Dodd}, which in the limit of an infinite 
system size $L\rightarrow\infty$ takes the form
\begin{equation}
\varphi^{kink}(x,t)=4\arctan\left\{\exp\left[\frac{x-Vt}{\sqrt{1-V^2}}\right]\right\}\;,
\label{kink_integr}
\end{equation}
where $V$ is the kink velocity, $|V|<1$. According to the energy balance equation~(\ref{en_bal}), 
in this case the internal energy current $J^{I}$ is the only
possible pathway to mediate an energy transport in the system. 
Using the expression~(\ref{kink_integr}) for the kink solution and the
definitions of the internal energy current $J^{I}$ (\ref{JINT}) and energy density 
$\rho$ (\ref{density}), it follows, that
the energy current associated with the kink motion can be obtained in full analogy with a moving 
point particle:
\begin{equation}
J^{I}\equiv V W^{kink}\;.
\label{Jintegr}
\end{equation}
Here the kink energy $W^{kink}$ is obtained from the energy density $\rho(\varphi)$:
\begin{equation}
W^{kink}=\int_0^L \rho[\varphi^{kink}] dx\;,\;
W^{kink}_{L\rightarrow\infty} = \frac{8}{\sqrt{1-V^2}}
\;.
\end{equation}

In the general case, when both $E$ and $\alpha$ are non-zero, a moving kink solution of Eq.~(\ref{sgE}) can 
persist as an attractor of the system with all the parameters (including the kink velocity) 
determined by the choice of $E$,  $\alpha$
and $L$. 
Our goal is to compute the energy current generated by such a moving kink. 
As will be shown below, an inhomogeneous exchange of energy
between the kink and the external degrees of freedom (force and dissipation), 
opens an additional path for energy transport in the system, which, 
together with the above internal current, 
constitute the full energy current balance generated by the moving kink.

\section{Constant driving force}
\label{sec_const}

Let us start with the seemingly simple case of the perturbed sin-Gordon equation (\ref{sgE}) -- 
the case of a constant driving force $E(t)=E\equiv Const.$ 
A kink moves in this case with a constant velocity $V$ (defined by a choice of the force strength $E$, as well as the damping constant $\alpha$ and the system length $L$), so that the corresponding attractor solution depends only on a single variable $\xi=x-Vt$ \cite{Dodd}: 
\begin{equation}
\label{MFRAME}
\varphi(x,t)=\psi(x-Vt)\equiv\psi(\xi).
\end{equation}

Due to the time homogeneity the total energy of the system $W$,
\begin{equation}
W=\int_0^L \rho[\varphi(x,t)]dx\;,
\label{energy_total}
\end{equation}
remains constant in time on the attractor solution. 
For a point particle the current balance would not change, since the energy loss due to
the damping would be compensated by the external dc field.
For the spatially extended kink this will not be the case anymore.

Let us assume, that the system size $L$
is chosen to be large enough, 
so that far away from the kink center 
the spatial field distribution asymptotically approaches the 
homogeneous ground 
state:
$\psi(\xi\to 0)\to\phi^v,\;\psi(\xi\to L)\to\phi^v+Q$. 
Note, that the external force $E$ shifts the degenerate ground state of the system from 
$\varphi_{G}^{(0)}=2\pi m, m=0,\pm1,\pm2,...$ to 
$\varphi_{G}^{(E)}=\varphi_{G}^{(0)}+\varphi^v(E),\;\varphi^v(E)\equiv\arcsin(E)$.

We denote  by $w(x,t)$ the amount of energy stored on the stripe $[0,x]$ at time $t$.
We further assume
that $x=0$ is far from the kink center so that the energy current density
vanishes there. Then
we can define the local energy current density
$j(x,t)$ as
\begin{equation}
j(x,t)=-\frac{\partial w(x,t)}{\partial t}=-\int_0^{x} \rho_t(x^{\prime},t) dx^{\prime}\;.
\label{jloc}
\end{equation}
Using the attractor property~(\ref{MFRAME}), we obtain:
\begin{equation}
j(x-Vt)\equiv j(\xi)=V\left[\rho(\xi)-\rho(0)\right]\;,
\label{jloc2}
\end{equation}
and the total energy current $J$ is thus given by
\begin{equation}
J=\int_0^{L} j(\xi) d\xi \equiv V\left[ \int_0^{L} \rho(\xi)d\xi - L\rho(0)\right]\;.
\label{JTOT_const}
\end{equation}
The quantity in the brackets in the r.h.s. of Eq.~(\ref{JTOT_const}) is the 
difference between the system energy with and without the kink. 
Indeed, in the presence of external force and damping, the ground state energy has shifted 
from zero to $W_0=L\cdot\rho(\phi^v)= L\cdot\rho(x=0)$. This 
difference is precisely the kink energy:
$W^k=W-W_0$. Thus, we arrive to the simple relation between the total energy current $J$, 
kink velocity $V$ and kink energy $W^k$:
\begin{equation}
J=V W^k\;.
\label{J_stac2}
\end{equation}
As expected, this expression corresponds to that of a point particle moving
with velocity $V$ and carrying energy $W^k$.
However, opposite to  
the integrable case $E=\alpha=0$, 
the internal current $J^I$ does not coincide with the total current 
$J$: 
$J^I\ne J$. Indeed, using the definition of the total current $J$ given in 
Eqs.~(\ref{jloc}),(\ref{JTOT_const}) and the energy balance equation~(\ref{en_bal}), 
we arrive at
\begin{eqnarray}
J= {J}^{I}+{J}^{E}\;,
\label{cur_bal}
\end{eqnarray}
where the \emph{exchange current} $J^E$ \cite{we} has been introduced
\begin{eqnarray}
\label{JEXCH_const}
J^{E}&=&V\int_{0}^{L}dx\int_{0}^{x}d\xi 
\left[\alpha V \left(\psi^{\prime}(\xi)\right)^2 + E \psi^{\prime}(\xi)\right]\;.
\end{eqnarray}

Following the definition of the internal current $J^I$ in Eq.~(\ref{JINT}), we obtain
\begin{eqnarray}
\label{JINT_const}
J^{I}&=&V\int_{0}^{L} \left(\psi^{\prime}(x)\right)^2 dx \;,
\end{eqnarray}
so that all components of the current balance equation~(\ref{cur_bal}) are 
expressed through the field $\psi$. Below we will calculate numerically 
the moving kink attractor solution $\psi(\xi)$ in order to estimate all the three 
currents for different values of the model parameters. 
Note, that according to Eq.~(\ref{JODE}) one can obtain the exact value for the 
internal current $J^I$ on the attractor solution in an independent way:
\begin{equation}
J_{exact}^{I}=-\frac{EQ}{\alpha}\;.
\label{JINT_exact}
\end{equation}
We will use this expression in order to control the accuracy of our numerical schemes.

The fact that the moving kink solution $\phi(x,t)$ depends only on the single 
variable $\xi=x-Vt$ allows to reduce the original field equation (\ref{sgE}) to the 
well studied driven and damped pendulum problem.
Indeed, substituting ansatz (\ref{MFRAME}) into Eq.(\ref{sgE}), we arrive to the 
following ODE for the function $\psi(\xi)$:
\begin{equation}
\label{ODE}
(1-V^2)\psi^{\prime\prime}=-\alpha V \psi^{\prime}+\sin(\psi)-E,
\end{equation}
which describes the evolution of an effective pendulum with the momentum of inertia $M=(1-V^2)$. 
The damping constant
$\gamma=\alpha V$ depends on the kink velocity $V$. The external force $E$ plays the 
role of an applied constant torque to the pendulum. By introducing the effective time
\begin{equation}
\label{time}
\tau=\frac{\xi}{\sqrt{1-V^2}},
\end{equation}
Eq.(\ref{ODE}) can be finally re-written for the function $y(\tau)\equiv\psi(\xi/\sqrt{1-V^2})$:
\begin{equation}
\label{oscillator}
\ddot{y}=-G\dot{y}+\sin(y)-E,
\end{equation}
with the damping constant $G$ defined through the original kink parameters:
\begin{equation}
\label{GE}
G(E,L)=\frac{\alpha V(E,L)}{\sqrt{1-V^2(E,L)}}.
\end{equation}
Here we emphasize the implicit dependence of $G$ on the strength of the external force 
$E$ and the system size $L$, since the kink velocity $V$ is 
unambiguously defined by these parameters.

Thus, a moving kink solution (\ref{MFRAME}) of the original problem (\ref{sgE}) 
with the boundary conditions (\ref{PBC}) corresponds to a limit cycle solution 
$\hat{y}(\tau)$ of the driven and damped pendulum (\ref{oscillator}). On this
limit cycle the pendulum performs 
periodic rotations with the period $T$ defined through the original 
system length $L$ and the kink velocity $V$ as:
\begin{equation}
\label{period}
T=\frac{L}{\sqrt{1-V^2}}.
\end{equation}
Note also, that the limit of an infinite system size $L$ corresponds to a limit cycle
solution of the pendulum with an infinitely large period $T$.

The existence and the characteristics of the limit cycle in Eq.(\ref{oscillator}) depend 
on the choice of values of $G$ and $E$ \cite{Fal76,LHM78,BBM+82}.
For each given value of $E$ from the interval $|E|\le 1$ there is a critical value of 
$G=G_{cr}(E)$, below which the limit cycle co-exists together with a stable static pendulum state
(the latter corresponds to the homogeneous ground state in the kink tails). The period of  
the limit cycle tends to infinity as $G$ approaches $G_{cr}$ from below, 
while above the critical value of $G$ the limit cycle disappears. As a consequence, the 
value $G_{cr}(E)$ corresponds to the limit of an infinite size $L$ in the original problem 
(\ref{sgE}) with the corresponding drive $E$, while any lower value $G<G_{cr}$ is associated 
with a moving kink in a finite-size system. For small values of $|E|$ the critical value of 
the effective damping can be approximated as \cite{BBM+82}
\begin{equation}
\label{g_approx}
G_{cr} \approx \frac{\pi E}{4}, \qquad |E| \ll 1.
\end{equation}
By using relation (\ref{GE}) one can derive the corresponding approximation for 
the kink velocity in an infinite-size system at low power of the external drive:
\begin{equation}
\label{v_approx}
V \approx \frac{\pi E}{\sqrt{16\alpha^2+\pi^2E^2}}, \qquad |E|\ll 1.
\end{equation}

To find the limit cycle solution of (\ref{oscillator}) and the corresponding 
value of $G_{cr}$ at arbitrary values of $E$ we use the Newton method in phase 
space $(\dot{y},y)$, rewriting Eq.(\ref{oscillator}) as \cite{BBM+82}:
\begin{eqnarray}
\label{phase_space1}
z(y)&=&\dot{y},\\
\label{phase_space2}
\frac{dz}{dy}&=&-G+\frac{\sin(y)-E}{z},
\end{eqnarray}
and computing the map
\begin{eqnarray}
\label{map}
\left\{z(y_0),y_0\right\}&\rightarrow&\left\{z(y_0+2\pi),y_0+2\pi\right\}, \\
\label{yini}
y_0&=&\arcsin(E)\;.
\end{eqnarray}
For any $z_0\equiv z(y_0)$ a unique periodic solution of (\ref{phase_space1},\ref{phase_space2}) 
satisfying condition $z(y_0+2\pi)=z_0$
can be found by tuning the parameter $G$, provided that the 
sign of $z_0$ is chosen in accordance with the sign of $E$. The critical value $G_{cr}$ 
is obtained by taking $z_0=0$. Note, that the value of $z_0$ is proportional to the value 
of $\psi^{\prime}(\xi=0,L)$. For any finite system size $L$ it is nonzero due to exponentially 
decaying kink tails.

\begin{figure}
\includegraphics[angle=270, width=0.45\textwidth]{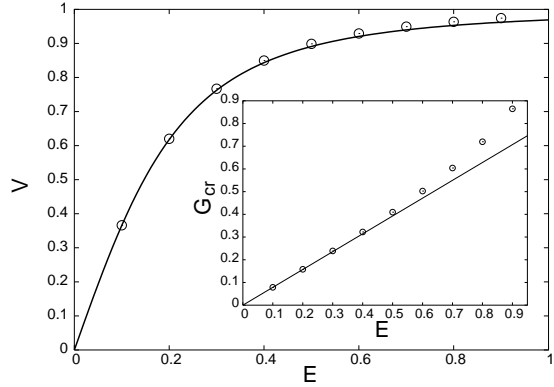}
\caption{Kink velocity as a function of the dc force $E$ for $\alpha=0.2$.
Open circles: numerical solution of (\ref{oscillator}), see text for details,
solid line: approximative solution (\ref{v_approx}).
Inset: critical value $G_{cr}$ from (\ref{GE}) for $\alpha=0.2$.
Open circles: numerical solution (see text for details), 
solid line: approximative solution (\ref{g_approx}).} 
\label{fig_osc1}
\end{figure}

In Fig.~\ref{fig_osc1} the computed values of $G_{cr}$, as well as the 
corresponding values of the kink velocity $V$, are shown with circles. 
In fact, the approximation for the kink velocity (\ref{v_approx}) works 
reasonably well even at values of $E$ close to unity, while the critical 
value $G_{cr}$ deviates stronger from the linear approximation (\ref{g_approx}) 
at values of the driving force close to $E=1$.

Once the limit cycle solution $\hat{y}(\tau)$ of (\ref{oscillator}) is found, 
the values of the total (\ref{JTOT_const}), exchange (\ref{JEXCH_const}) and 
internal (\ref{JINT_const}) currents for the corresponding moving kink solution can be obtained:
\begin{eqnarray}
\label{jexch_new}
J^{E}&=&\frac{\alpha G}{\alpha^2+G^2}\left(D-F\right),\\
\label{jint_new}
J^{I}&=&\frac{2G}{\alpha}F\\
\label{jtot_new}
J&=&\frac{G}{\alpha}\frac{\alpha^2+2G^2}{\alpha^2+G^2}F+\frac{\alpha G}{\alpha^2+G^2}D\\
\nonumber
 &\equiv& J^{E}+J^{I},
\end{eqnarray}
where the functions $F$ and $D$, are related to the mean pendulum kinetic and 
potential energies:
\begin{eqnarray}
\label{FE}
F(E,L)&=&\int_{0}^{T}d\tau \frac{\dot{y}^2}{2},\\
\label{DE}
D(E,L)&=&\int_{0}^{T}d\tau \left[\sqrt{1-E^2}-\cos(y)\right].
\end{eqnarray}
Here the implicit dependence on the original system length $L$ comes through the pendulum period $T$.

Therefore, the three functions $D$ (\ref{DE}), $F$ (\ref{FE}) and $G$ (\ref{GE}), 
uniquely defined for a limit cycle of (\ref{oscillator}) for a given value of the 
external field $E$ and $\dot{y}(0)$, determine all the characteristics of the 
corresponding moving kink problem: its spatial profile, velocity, energy, internal, 
exchange and total currents. In Table~\ref{tab1} we list the values of all three 
currents calculated for the damping constant $\alpha=0.2$ and different strength 
of the external driving force $E$ by means of the pendulum approach, corresponding 
to the limit of $L\rightarrow\infty$ in the original system (\ref{sgE}). They perfectly 
agree with the results of direct numerical integration of Eq.~(\ref{sgE}). In order 
to control the accuracy of our numerical scheme, we estimate the error of the calculated 
internal current $|\Delta J^I|$, for which we know the exact result $J^I_{exact}$ (\ref{JODE}). 
We also indicate in Table~\ref{tab1} the accuracy $\Delta J$ at which the current ballance 
equation~(\ref{cur_bal}) is fulfilled: $\Delta J=J-J^I-J^E$, and the relative strength of 
the exchange current $|J^E/J|$.

\begin{table}
\begin{tabular}{c|c|c|c|c|c|c}
$E$ & $J^I$  & $J^E$ & $J$  & $|J^E/J|$ & $|\Delta J^I|$ & $|\Delta J|$\\
\hline
$-0.05$ & $1.570796328$ & $-7.64938\cdot 10^{-4}$  & $1.570031391$  & $4.9\cdot 10^{-4}$ & $1.7\cdot 10^{-9}$ & $1.0 \cdot 10^{-10}$ \\
$-0.1$ &  $3.141592658$  & $-5.531719\cdot 10^{-3}$ & $3.313606094$  & $1.8 \cdot 10^{-3}$ & $4.8\cdot 10^{-9}$& $1.6 \cdot 10^{-10}$ \\
$-0.15$ & $4.71238899$  & $-1.610701 \cdot 10^{-2}$ & $4.69628198$  & $3.4 \cdot 10^{-3}$& $1.2 \cdot 10^{-8}$ & $1.0 \cdot 10^{-9}$\\
$-0.2$ &  $6.28318532$  & $-3.210284 \cdot 10^{-2}$ & $6.25108248$  & $5.1 \cdot 10^{-3}$ & $1.3 \cdot 10^{-8}$ & $1.0 \cdot 10^{-9}$\\
$-0.25$ & $7.8539826$  & $-5.21135 \cdot 10^{-2}$ & $7.8018691$  & $6.7 \cdot 10^{-3}$ & $9.4 \cdot 10^{-7}$ & $1.0 \cdot 10^{-8}$\\
$-0.3$ &  $9.4247783$ & $-7.50335 \cdot 10^{-2}$ & $9.3497448$ & $8.0 \cdot 10^{-3}$ & $3.7 \cdot 10^{-7}$ & $5.0 \cdot 10^{-8}$ \\
$-0.35$ & $10.995576$ & $-9.9883 \cdot 10^{-2}$ & $10.895692$ & $9.2 \cdot 10^{-3}$ & $1.2 \cdot 10^{-6}$  & $1.0 \cdot 10^{-7}$ \\
$-0.4$ & $12.566373$ & $-1.26308 \cdot 10^{-1}$ & $12.440065$ & $1.0 \cdot 10^{-2}$ & $2.4 \cdot 10^{-6}$ & $1.0 \cdot 10^{-7}$ \\
$-0.5$ & $15.70799$ & $-1.8326 \cdot 10^{-1}$ & $15.52472$ & $1.2 \cdot 10^{-2}$ & $2.2 \cdot 10^{-5}$ & $1.0 \cdot 10^{-6}$ \\
$-0.6$ & $18.84957$ & $-2.4832 \cdot 10^{-1}$ & $18.60126$ & $1.3 \cdot 10^{-2}$ & $2.2 \cdot 10^{-5}$ & $1.0 \cdot 10^{-6}$ \\
$-0.7$ & $21.99116$ & $-3.2617 \cdot 10^{-1}$ & $21.66500$ & $1.5 \cdot 10^{-2}$ & $2.0 \cdot 10^{-5}$ & $1.0 \cdot 10^{-6}$ \\
$-0.8$ & $25.1331$ & $-4.258 \cdot 10^{-1}$ & $24.7073$ & $1.7 \cdot 10^{-2}$ & $3.0 \cdot 10^{-4}$ & $1.0 \cdot 10^{-5}$ \\
\end{tabular}
\caption{The values of the internal, exchange and total 
currents for the damping constant $\alpha=0.2$ 
and an infinite system size $L\rightarrow\infty$, calculated within the pendulum approach . 
$\Delta J^I$ and $\Delta J$ 
are the errors of the calculated internal current 
and the accuracy at which the current balance relation (\ref{cur_bal}) is fulfilled, 
respectively (see the main body text for details).}
\label{tab1}
\end{table}

The surprising result of the presented analysis is that for the case of a constant 
driving force, when the kink is known to behave similar to a point particle
without any internal degrees of freedom being excited, the exchange current is nonzero! 
It is relatively weak, but both its absolute value and its relative strength are significantly 
larger than the estimated numerical error and increase with increasing driving force $E$. 

\begin{figure}
\includegraphics[angle=-90, width=0.5\textwidth]{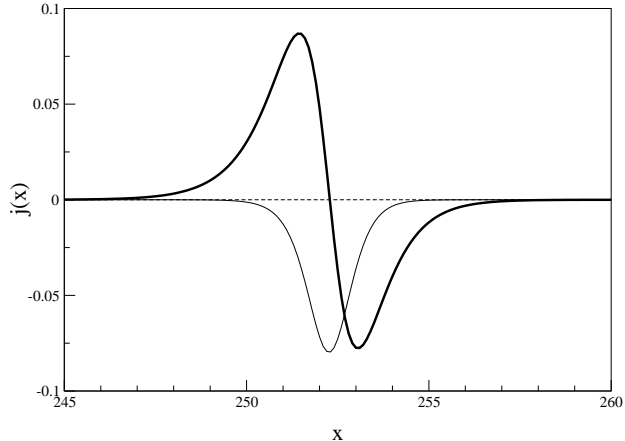}
\caption{ The densities of the scaled internal current, $0.02\; j^{I}(x)$ (thin
line), and of the exchange current, $j^{E}(x)$ (bold line), for the
$E=0.2$,  
$\alpha=0.2$ and $L=500$.
\label{fig_curdens}}
\end{figure}

Since for a constant driving force all currents are time-independent, 
we can easily define the densities of internal and exchange currents
\begin{eqnarray}
\label{JINT_dens}
j^{I}(x)&=&V \psi^{\prime}(x)^2 \;,\\
\label{JEXCH_dens}
j^{E}(x)&=&V\int_{0}^{x}d\xi 
\left[\alpha V \left(\psi^{\prime}(\xi)\right)^2 + E \psi^{\prime}(\xi)\right]\;,
\end{eqnarray}
using Eqs.~(\ref{JINT_const}) and (\ref{JEXCH_const}). The corresponding profiles are plotted 
in Fig.~\ref{fig_curdens} for the case $\alpha=0.2,\;E=0.2$.
First it is evident from (\ref{JEXCH_dens}), that the exchange current
density is nonzero, similar to the internal current density (\ref{JINT_dens}).
This is due to the finite spatial extent of the kink, which
causes a corresponding spatially (and thus temporally as well)
inhomogeneous energy exchange between the field $\varphi$, the
external forcing $E$ and the damping term.
A total vanishing of $J^E$ could only hold due to some spatial symmetry
of $j^E(x)$ which in turn has to be generated by some symmetry of the kink profile.
However the kink profile in the presence of driving force and damping is asymmetric.
That is the main reason for the nonvanishing exchange current contribution.

Finally, we note, that the pendulum approach is valid only for the case of $|E|\le 1$. Indeed, 
in the case of $|E|> 1$ the limit cycle of (\ref{oscillator}) persist, but the stable fixed 
point disappears. This means that one no longer has a homogeneous ground state in the kink tails. 
In this case the ansatz (\ref{MFRAME}) is not valid and the pendulum approach 
becomes inappropriate. Similarly, when $|E|> 1$ the proposed scheme for the calculation of the 
total and exchange currents, Eqs.~(\ref{JTOT_const}) and (\ref{JEXCH_const}), breaks, since 
the homogeneous ground state in the kink tails is no longer supported by the starting equations.

\section{Time-periodic driving force: soliton ratchets}
\label{sec_rat}

\begin{figure}
\includegraphics[angle=270,width=0.7\textwidth]{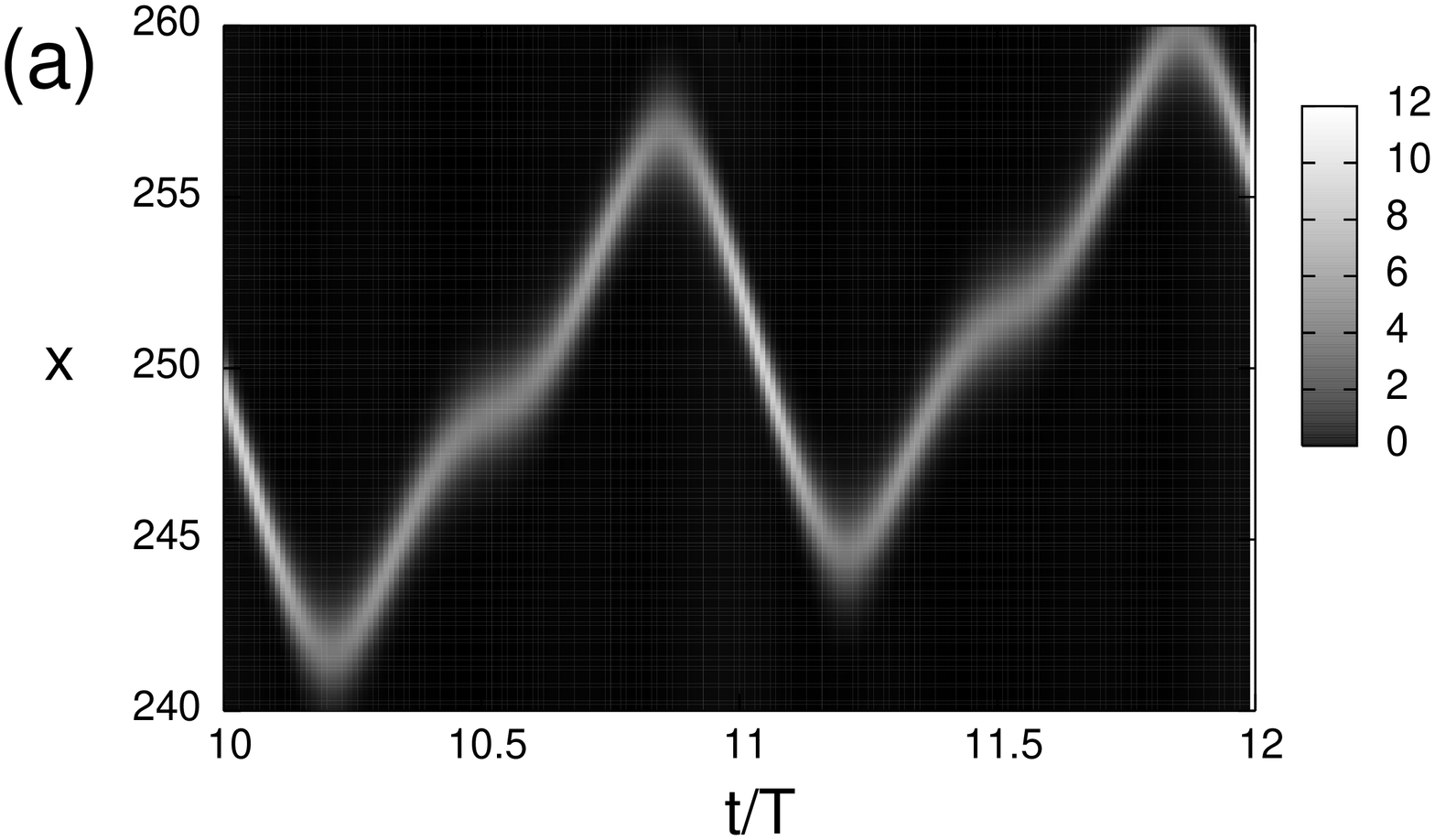}
\includegraphics[angle=270,width=0.7\textwidth]{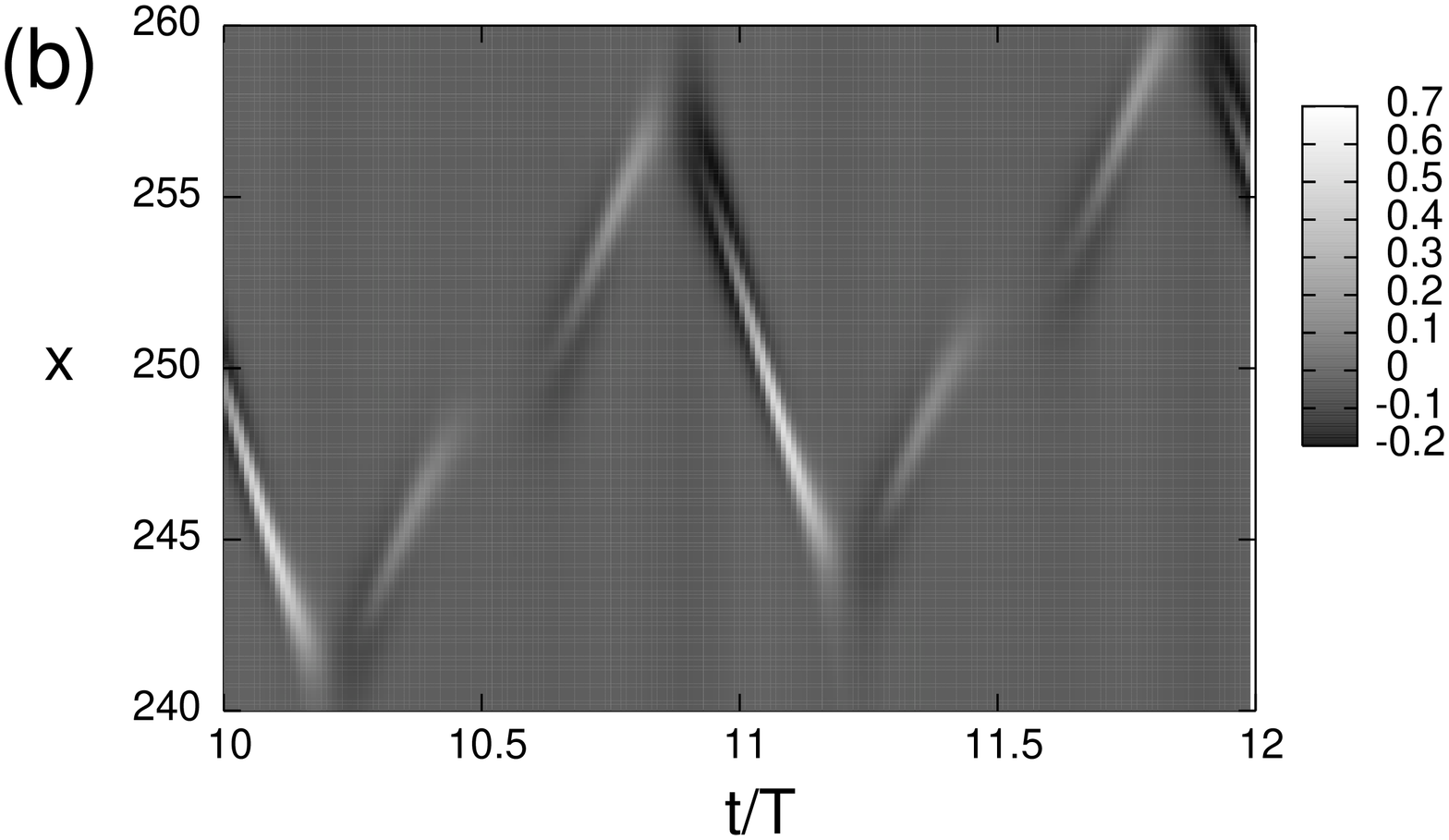}
\caption{ Space-time evolution of the soliton ratchet
with the $E_1=E_1=0.2,\;\omega=0.1,\;\Theta=0$ and
$\alpha=0.2,\;L=500$:
(a) Contour plot of the energy density (\ref{density});
(b) Contour plot of the function
$\phi(x,t)$ defined in Eq. (\ref{phi}).
}
\label{fig_ratchet}
\end{figure}

In this section we proceed to the case of a time-periodic driving force 
$E(t+T)=E(t)$ with zero mean value. 
For the case of an AC driving force, which possesses the time-shift symmetry
\begin{eqnarray}
E(t)= -E(t+T/2)\;, 
\label{time-shift}
\end{eqnarray}
the combined symmetry transformation
\begin{eqnarray}
x \rightarrow -x, ~~ \varphi \rightarrow -\varphi + Q, ~~ t
\rightarrow t + \frac{T}{2}
\label{sym-transf}
\end{eqnarray}
leaves Eq.~(\ref{sgE}) invariant while changing the sign of the kink velocity $V$ 
defined e.g. as \cite{Flach, Salerno_Zol}
\begin{eqnarray}
V(t)=\frac{1}{Q}\int^{L}_{0} x \varphi_{tx}dx\;.
\label{kink_vel}
\end{eqnarray}
If only one kink attractor persists, then on this attractor the kink will not
move, but oscillate at the best \cite{Flach,we}. This is the typical situation observed
in various numerical simulations.
In the unlikely case that a kink attractor persists on which the kink is moving with a nonzero average
velocity, then by symmetry another attractor exists as well, on which the kink
moves with the opposite velocity.
However, the above time-shift symmetry (\ref{time-shift}) is generally removed by choosing
the bi-harmonic driving force
\begin{eqnarray}
E(t)= E_{1}\cos[\omega (t-t_0)]+ E_{2}\cos[2 \omega (t-t_0) + \Theta]\;,\;
\omega=\frac{2\pi}{T}\;,0<\Theta<2\pi\;.
\label{ratchet_force}
\end{eqnarray}
As a consequence, the kink on the attractor solution will in general have
a nonzero velocity   (provided no other hidden symmetries of 
the system persist). This results in a unidirectional (in average) motion of the kink under the influence 
of zero-mean periodical driving force given in 
Eq. (\ref{ratchet_force}) \cite{Flach, Salerno_Zol, we}. 
The corresponding kink ratchet effect was observed both in numerical simulations \cite{Flach, Willis} 
and in experiments with an annular Josephson junction \cite{Ustinov}. In Fig.~\ref{fig_ratchet}(a)  
the characteristic evolution of the energy density~(\ref{density}) in the kink ratchet dynamics 
with the bi-harmonic driving force
(\ref{ratchet_force}) is shown. A gradual drift of the energy density peak, associated with the 
kink center, is clearly observed.

The 
unidirectional motion of the kink should result in the appearance of a non-zero averaged 
energy current. Simultaneously, as follows from Eq.(\ref{JODE}), for any zero-mean $E(t)$,
the time average value of the internal current $J^I$ should be zero on the attractor 
\cite{Joergensen, Olsen, Quintero}. Therefore, 
the internal current does not contribute to the energy transport associated with the kink motion. 
This is impossible from the point of view of a point particle,
since then the 
internal current $J^I$ corresponds to an effective particle momentum \cite{Dodd, Joergensen, Olsen, Bonilla}, and the 
kink will perform a momentumless motion in the ratchet case. This circumstance
raised discussions of whether the observed rectification of the kink motion
is due to additional white noise or discretization effects (see e.g. \cite{Quintero}). 
However, such a momentumless motion does not contradict the energy balance 
principles, since there exists an alternative path for the energy flow in the 
system -- the exchange current $J^E$, as we demonstrated in the previous section. In the case 
of bi-harmonic driving force all the energy is transported through the exchange current, so 
that the time-averaged current balance equation~(\ref{cur_bal}) 
reads:
\begin{eqnarray}
\mathcal{J}=\mathcal{J}^{E}, ~~~ \mathcal{J}^{I}=0\;.
\label{cur_bal_ratchet}
\end{eqnarray}
Here we use calligraphic letters to indicate 
time-averaged quantities:  $\mathcal{J}\equiv(1/T)^2\cdot\int_0^T dt \int_t^{t+T}dt_1J(t_1)$
(details of the averaging procedure will be discussed below, see also Ref.~\cite{we}).

Our goal now is to estimate the energy current $\mathcal{J}$ generated by a moving kink. 
Taking the system size $L$ much larger than the characteristic
oscillation distance $L_p$ which the moving kink covers during a single period of the external force, 
we may assume
that far from the kink center the field distribution in space is homogeneous.
We separate the field variable 
$\varphi(x,t)$ into a localized kink part,
$\varphi^{k}(x,t)$, where $\varphi^{k}(x\rightarrow 0,t)=
0;~\varphi^{k}(x\rightarrow L,t)= Q$, and a background ({\it vacuum})
part $\varphi^{v}(t)$ which depends only on time \cite{Olsen}:
\begin{eqnarray}
\varphi(x,t)= \varphi^{k}(x,t)+\varphi^{v}(t)\;.
\label{field_sep}
\end{eqnarray}
The vacuum part must satisfy Eq.(\ref{sgE}) in the absence of the kink, i.e. when $Q=0$.
Therefore it does not contribute to any energy
transport \cite{Flach}.

On the attractor the dynamics of the system (\ref{sgE}) is given by
\begin{eqnarray}
\varphi^k(x,t+T)=\varphi^k(x-\mathcal{V}T,t)\;,
\;
\varphi^{v}(t+T)=\varphi^{v}(t)\;,
\label{attractor}
\end{eqnarray}
where $\mathcal{V}$ is the averaged kink velocity. 
Note that all integral system characteristics, such as the
total energy of the system, the kink velocity, the
energy current, etc. are periodic functions of time
with the period $T$.

Since the total system energy $W$ (\ref{energy_total}) is not conserved, we can not
introduce
instantaneous energy currents. However, since $W$ is a time-periodic function on 
the kink attractor solution, 
we estimate currents by considering energy density distribution changes over the period of attractor, 
when the total system energy is restored.
Assuming that $x=0$ corresponds to a kink tail point, the 
amount of energy $\triangle w(x,t)$:
\begin{eqnarray}
\nonumber
\triangle w(x,t)= \int_{0}^{x} [\rho (x',t) - \rho (x',t+T)]
dx' =  \\
-\int_{0}^{x} dx' \int_{t}^{t+T} \rho_{t} (x',t') dt'\;
\label{dw}
\end{eqnarray}
which the part of the system $[0,x]$ looses or gains during one period $T$ has to
be transported through 
the point $x$ by
the energy current $j(x,t)=\triangle w(x,t)/T$. 
Both quantities, $j(x,t)$ and $\triangle w(x,t)$, are obviously periodically dependent on the reference time $t$.
However, it is missleading to associate them with any instantaneous characteristics of the system,
as $\triangle w(x,t)$ is defined on the timescale of the attractor period $T$.
In order to eliminate the dependence on reference time $t$ we average both quantities over the period $T$.
Finally, the total averaged energy current in the system is obtained by the additional
integration over the space: 
\begin{eqnarray}
\mathcal{J}=\frac{1}{T} \int_0^L dx \int_0^T dt \; j(x,t) \;\;,
\;\;
j(x,t)=\frac{1}{T} \triangle w(x,t) \;.
\label{totalcurrent}
\end{eqnarray}

Using Eqs.~(\ref{sgE}), (\ref{attractor}) and  the multiplicative integration
rule \cite{Ryzhik} we finally obtain (see Appendix for details):
\begin{eqnarray}
\mathcal{J}=\mathcal{V}  \langle \int_{0}^{L} \rho[\varphi(x,t)] dx -
L  \rho[\varphi^{v}(t)] \rangle_{T}
\label{JTOT}
\end{eqnarray}
with $\langle ... \rangle_T = 1/T \int_0^T ... dt$. As before, 
the difference between the total system energy $W$ and the background energy $W_0$,  
averaged in the r.h.s. of Eq.~(\ref{JTOT}), denotes the kink energy $W^k$. Therefore, the
general result for the total energy current, given in Eq.~(\ref{JTOT_const}), still holds 
in this case, if \emph{using averaged values}. 

Using the results (\ref{dw},\ref{totalcurrent}) we can now derive an expression for the 
averaged exchange current $\mathcal{J}^{E}$ 
from the energy balance equation (\ref{en_bal}):
\begin{eqnarray}
\label{JEXCH}
\mathcal{J}^E &=& \frac{1}{T} \int_0^T dt \int_0^L dx j^E(x,t)\;,\\
j^{E}(x,t)&=& \frac{1}{T} \int_{t}^{t+T} dt' \int_{0}^{x} dx'
\phi(x',t') \;,
\\
\phi(x,t)&=&\phi[ \varphi ]= \alpha \varphi_{t}^{2} -
E(t)\varphi_{t}\;.
\label{phi}
\end{eqnarray}

Note, that similar to previously defined $j(x,t)$ in Eq.~(\ref{JTOT}), 
the quantity $j^E(x,t)$ can not be treated as an instantaneous
current density value.

Again, using the multiplicative integration rule \cite{Ryzhik}, Eqs.~(\ref{JEXCH})-(\ref{phi}) can be reduced 
to the more compact form (see Appendix for details):
\begin{equation}
\label{JEXCH_red}
\mathcal{J}^E=- \langle \int_{0}^{L} \left[ \phi(x-\mathcal{V}T,t)-\phi^v(t)\right]\cdot\left[x-\mathcal{V}t\right] dx
\rangle_{T} \;,
\end{equation}
where $\phi^v(t)\equiv\phi[\varphi^v(t)]$. 

The appearance of a non-zero exchange current in the system is mediated by a spatially
and temporally inhomogeneous 
energy exchange between the moving kink and the external degrees of freedom.
The information about the energy exchange process is contained in the  
function $\phi(x,t)$~(\ref{phi}) introduced above. 
In Fig.\ref{fig_ratchet}(b) the space-time evolution of $\phi$ 
is plotted. The energy is exchanged and transported
in a cyclic way: first the kink absorbs energy in its rare tail, then it releases energy in 
its front, then it absorbs energy in its front and
finally releases energy in the rare tail. 

\begin{figure}
\includegraphics[angle=0, width=0.45\linewidth]{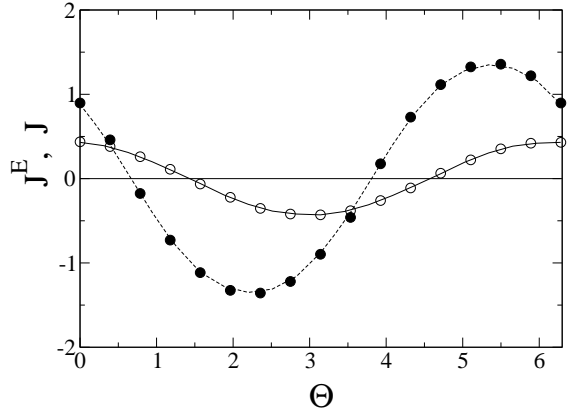}
\caption{ 
Numerically computed average values of the exchange current $\mathcal{J}^{E}$ (lines) and 
total current $\mathcal{J}$ (circles) as a function of the driving force parameter $\Theta$
for $\alpha=0.2$ (solid line, filled circles) and $\alpha=0.05$ (dashed
line, open circles). Other parameters are the same as in Fig.~\ref{fig_ratchet}.
}
\label{fig_cur}
\end{figure}

The numerical simulations of the kink dynamics~(\ref{sgE}) confirm that the balance between 
averaged values of the total and exchange currents holds, see Fig.~\ref{fig_cur}. 
The absolute values of the exchange current increased, as compared to the case 
of a constant driving force (cf. with Table~\ref{tab1}). 
It may be due to the excitation of kink shape modes, which are clearly observed in 
Fig.~\ref{fig_ratchet} -- the kink is much more compressed when moving opposite to its 
average propagation direction as compared to the times when it moves in the same direction.

Note that the rectification effect vanishes at a certain value of the phase mismatch $\Theta$ 
between the two components of the driving force. This value depends on the damping constant 
$\alpha$. That is due to another symmetry property of the system~(\ref{sgE}) in the 
Hamiltonian limit $\alpha\rightarrow 0$. Indeed, in that limit the system possesses the time-reversal 
symmetry $t\to -t$. That symmetry operation
changes the sign of the kink velocity~(\ref{kink_vel}), provided the driving 
force is taken to be symmetric in time: $E(-t)=E(t)$.
Therefore, in the Hamiltonian limit the rectification effect should disappear at 
$\Theta=0,\pi$. In 
the underdamped regime $\alpha \ll 1$ this effect persists at some value of $\Theta$ close to 
$\Theta=0,\pi$ \cite{Flach}.

If a DC component is added to the bi-harmonic driving force~(\ref{ratchet_force}), 
a careful tuning of the parameters can lead to an exact cancellation
of both DC and AC force components, so that $\mathcal{V}=0$.
Then the averaged total current $\mathcal{J}$ (\ref{JTOT}) is exactly zero.
However, according to Eq. (\ref{JINT_exact}), the internal current is 
$\mathcal{J}^{I} =-QE^{stop}/\alpha$, where $E^{stop}$ is the DC component of the driving force.
This implies an exact
balance between the two nonvanishing current components:
\begin{eqnarray}
\mathcal{J}=0,~~ \mathcal{J}^{E} = -\mathcal{J}^{I}\;.
\label{curbal_acdc}
\end{eqnarray}
For example, taking parameter values from Fig.~\ref{fig_ratchet}, 
we found $E^{stop}\approx 0.01705$ and $\mathcal{J}^{E}=-\mathcal{J}^{I}\approx 0.536$.
Note also, that for any $0<E<E^{stop}$
the
sign of the total momentum of the system $\mathcal{J}^{I}$ is
{\it opposite} to the sign of the kink velocity $\mathcal{V}$,
and the internal current is pumping energy in a direction opposite to the kink motion.

\section{Effects of spatial discretization}
\label{sec_discr}

Up to now we discussed the energy flow due to kink motion in a spatially extended \emph{continuous} system.
It is known, that the spatial discreteness of a system can drastically change 
the soliton dynamics \cite{Braun_Rep}.
By breaking translational invariance of the system, discreteness induces a Peierls-Nabarro potential 
which strongly influences the translational motion of the kink \cite{Braun_Rep}. It is also known, that
additional internal modes of the  kink may appear in discrete systems \cite{Braun}.
Their existence can essentially modify energy exchange mechanisms between the moving kink and external 
driving forces and damping. As a consequence, the relative contributions 
of the internal and exchange currents to the total energy current can be strongly 
changed in the discrete case both
for DC and AC driving forces.

Let us consider the discrete version of Eq.~(\ref{sgE}) also known in literature 
as the damped and driven Frenkel-Kontorova chain \cite{Frenkel}:
\begin{eqnarray}
\ddot{\varphi}_n-C^2\left(\varphi_{n+1}+\varphi_{n-1}-2\varphi_n\right)=-\alpha\dot{\varphi}_n-
\sin \varphi_n+E(t)\;.
\label{FK}
\end{eqnarray}
It has been used in solid state theory for modeling dislocation dynamics in crystals. 
This equation also describes fluxon dynamics in one-dimensional arrays of coupled Josephson 
junctions \cite{Ustinov_D}.
It is important to note, that Eq.~(\ref{FK}) is often used for numerical simulations of the 
original continuous 
sin-Gordon system (\ref{sgE}), where $C\sim1/h$ is the inverse mesh size or the lattice spacing 
(i.e. $C\to \infty$ corresponds to the continuous limit). 
In this respect it is also important to investigate
possible effects of discretization which might appear during the numerical calculation of energy 
currents of an underlying spatially continuous system.

In order to relate the above results for the continuous sin-Gordon model~(\ref{sgE}) 
to the discrete system~(\ref{FK}), 
we use the transformation $x\to n/C,\;\varphi(x,t)\to\varphi(n/C,t)=\varphi_n(t)$. 
Thus, all spatial derivatives should be replaced by finite differences weighted by the factor $1/h=C$, while
all spatial integrals should be replaced by discrete sums over lattice sites with the weight factor $1/C$.
With that we obtain the following discretized versions for the internal~(\ref{JINT}) 
and total~(\ref{JTOT}) 
energy currents:
\begin{eqnarray}
\label{JINT_D}
J^I&=&\frac{1}{C}\sum_n j^I_n\;,\\
\label{JTOT_D}
J&=&V_d\sum_n \left[ \rho_n-N\rho^v\right],
\end{eqnarray}
where $V_d$ is the kink velocity calculated in units of lattice site differences per time. 
The local internal current $j^I_n$ and energy density $\rho_n$ are yet to be defined.

We note, that the Frenkel-Kontorova model~(\ref{FK}) corresponds to the Hamiltonian
\begin{equation}
H=\sum_n \rho_n=\sum_n\left\{\frac{\dot{\varphi}_n^{2}}{2} + \frac{C^2}{4}\left[
\left(\varphi_{n+1}-\varphi_n\right)^2+\left(\varphi_{n}-\varphi_{n-1}\right)^2
\right]
+1-\cos (\varphi_n)\;.
\right\}
\end{equation}

\begin{figure}
\includegraphics[angle=0,width=0.8\linewidth]{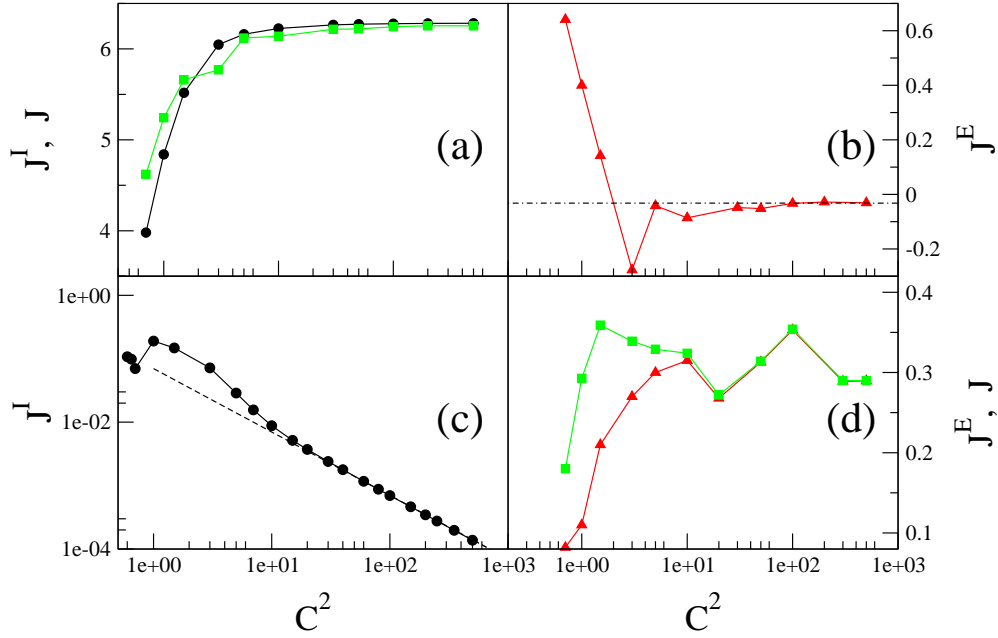}
\caption{
The internal (filled circles), the exchange (triangles) and the total (squares) energy currents 
as functions of the discretization parameter $C^2$.
(a) and (b): the case of the DC external force $E=0.2$;
(c) and (d):  the case of the 
biharmonic driving force~(\ref{ratchet_force}) with $E_1=E_2=0.2$ and $\Theta=0$. 
$\alpha=0.2$ for
all cases. The dash-dotted line in (b) indicates the exchange current value, $J^{E} \approx -0.0320128$,
calculated for the continuum limit within the pendulum approach (see Table 1.). The dashed line in (c)
corresponds to the power-law asymptotic  $J \sim C^{-2}$.
}
\label{fig_discr}
\end{figure}

Following essentially the same steps as in Sec.~\ref{sec_model}, we can derive the discrete version of the
energy balance equation~(\ref{en_bal}) for the discrete energy density $\rho_n$ 
(energy of an individual lattice site):
\begin{eqnarray}
\label{en_bal_d}
\frac{d\rho_n}{dt}&=&-C\left(j^I_n-j^I_{n-1}\right)-\alpha \dot{\varphi}_n^2+E(t)\dot{\varphi}_n\;,\\
\label{jh_d}
j^I_n&=&\frac{C}{2}\left(\dot{\varphi}_{n+1}+\dot{\varphi}_{n}\right)
\left(\varphi_{n+1}-\varphi_{n}\right)\;.
\end{eqnarray}
The internal current density $j^I_n$ (also known as the local heat flux \cite{Lepri}) 
is the discrete analogue of the internal current density~(\ref{JINT}) of the continuum case.

In contrast to the continuum limit, it is not straightforward to define a discrete version of the exchange 
current on the basis of the last two terms in the r.h.s. 
of the balance equation~(\ref{en_bal_d}). However, the exchange current can be computed via the difference 
between the total and internal currents, $J^{E}=J-J^{I}$.

In Fig.~\ref{fig_discr} the dependencies of the computed energy currents on the discretization 
parameter $C$ are plotted for 
the cases of DC and AC external driving forces. Increasing $C$, in both cases all the currents 
clearly converge to the corresponding values 
obtained in the continuum limit: in the DC case the dominant energy pathway is realized through 
the internal current, 
while in the AC case the only remaining path for energy transport is mediated by the exchange current.
In the latter case the internal current scales down to zero as $J^I\sim C^{-2}$ for $C^2\gtrsim 10$, 
which follows from the approximation of integrals by discrete sums \cite{Ryzhik}.
Spatial discretization can drastically change the ratio between the two currents 
when $C^2 \lesssim 10$ both in the case 
of a constant external bias, and in
the case of a soliton ratchet \cite{Flach}. Since for these values of $C$ the Peierls-Nabarro potential
is still exponentially small \cite{PN}, the cause of the corrections
is simply the change of the kink shape which is of order $1/C^2$ \cite{Flach_Kladko}.

\section{Conclusions}
\label{sec_fin}

We discussed the issue of soliton-assisted energy transport in 
spatially extended systems with external driving and damping. 
Considering the topological soliton (kink) motion in the well-known sin-Gordon model, 
we showed, that the conventional description of energy transport based on the total field momentum 
does not provide one with the correct value of energy flux in the system. In particular, in the 
case of a directed
soliton motion under the influence of time-periodic external forces with zero mean, 
the averaged value of the total 
field momentum is known to be 
strictly zero \cite{Joergensen, Olsen, Quintero, we}, while the energy transport associated with the soliton motion 
is obviously non-zero. 
We identified a new energy pathway -- \emph{the exchange current} -- which is entirely
mediated by the spatial and temporal inhomogeneity of the system state.
Even for the case of a DC external force, the exchange
current is found to be small but nonzero. 
Combining both DC and AC driving we obtain situations when the total field momentum
is nonzero but the kink does not move on average or moves even in the direction
opposite to the field momentum.

The approach to the energy transport in spatially extended systems which is based 
solely on the consideration of the 
total field momentum reduces the kink motion to that of a point particle.
However, the soliton motion in presence 
of AC driving forces is always
accompanied by the excitation of internal modes \cite{Luis1}, which 
makes the dynamical properties of the moving soliton essentially different
from those of a point particle. 
Although we stress here that 
even in cases where no shape modes are excited
the total field momentum is not sufficient to obtain the total energy current, see Sec.~\ref{sec_const}.
It is the spatial extent of the soliton which makes the difference
for a damped and driven moving soliton.

The relative contributions of the two current components, the internal and the exchange one, 
can change when considering spatially discrete systems, 
see Sec.~\ref{sec_discr}.
In particular, for the soliton ratchet case,
discreteness induced corrections to the internal current 
invalidate Eq.~(\ref{JODE}), so that
$J^{I}$  also contributes to the total energy flow. 
This explains earlier obtained numerical results from Ref.~\cite{Flach}.

Finally, we would like to note, that our results are also instructive for the
general case of spatially extended systems coupled to external
driving forces or other degrees of freedom, see e.g. Refs.~\cite{Barashenkov, Sukstanskii}.
We also mention that the damped and driven sin-Gordon system in Eq.(\ref{sgE}), as well as its 
discretized version in Eq.~(\ref{FK}), 
are relevant 
physical models of annular \cite{Ustinov, Goldobin} and coupled \cite{Ustinov_D} Josephson junction 
oscillators, respectively. 
Therefore, an intriguing question rises about the possibility of detection of the new exchange 
current mechanism on the basis of  available experimental data (current-voltage characteristics, 
spectra of emitted radiation, etc). Indeed, since directed soliton motion is observed in such
experiments, the internal and exchange currents have a clear physical meaning, though probably
not easy to measure. 

\vspace{1cm}

\appendix{\bf APPENDIX}
\label{sec_app}

Using Eq.~(\ref{dw}), let us rewrite the expression~(\ref{JTOT}) for the energy current $\mathcal{J}$ in the following form:
\begin{equation}
\label{A1}
\mathcal{J}=\frac{1}{T^2}\int_0^Tdt\int_0^Ldx\int_0^x dx^{\prime} \left[\rho(x^{\prime},t)-\rho(x^{\prime},t+T)\right] \;.
\end{equation}
According to the multiplicative integration rule \cite{Ryzhik}
\begin{equation}
\int_0^Ldx\int_0^x dx^{\prime} \left[\rho(x^{\prime},t)-\rho(x^{\prime},t+T)\right]=
\int_0^L (L-x)\left[\rho(x,t)-\rho(x,t+T)\right] dx\;.
\label{A2}
\end{equation}

Since the total energy of the system $W$ is a periodic function with period $T$, the integral 
$\int_0^L L\left[\rho(x,t)-\rho(x,t+T)\right]dx=L\left[W(t)-W(t+T)\right]\equiv 0$. Thus, expression~(\ref{A1}) is reduced to
\begin{equation}
\label{A3}
\mathcal{J}=\frac{1}{T^2}\int_0^Tdt\int_0^L x \left[\rho(x,t)-\rho(x,t+T)\right] dx \;.
\end{equation}
Let us introduce the energy-weighted center of kink
\begin{equation}
\hat{X}(t)=\frac{\int_0^L x \left[\rho(x,t)-\rho^v(t)\right] dx}{\int_0^L\left[\rho(x,t)-\rho^v(t)\right] dx}\;,
\label{A4}
\end{equation}
where $\rho^v(t)\equiv\rho[\varphi^v(t)]$. Then we obtain
\begin{equation}
\mathcal{J}=\frac{1}{T^2}\int_0^Tdt\int_0^L \left[\rho(x,t)-\rho^v(t)\right]\cdot\left[\hat{X}(t+T)-\hat{X}(t)\right] dx \;.
\end{equation}
Since $\left[\hat{X}(t+T)-\hat{X}(t)\right]=\mathcal{V}T$, finally we arrive to the following expression for the total energy current:
\begin{equation}
\mathcal{J}=\frac{\mathcal{V}}{T} \int_0^T dt \int_{0}^{L} \left[\rho(x,t)-\rho^v(t)\right] dx
\label{A5}
\end{equation}

In a similar way one can derive the compact expression~(\ref{JEXCH_red}) for the exchange current. We re-write the expression~(\ref{JEXCH}) in the following form:
\begin{equation}
\label{A6}
\mathcal{J}^E=\frac{1}{T^2}\int_0^Ldx\int_0^x dx^{\prime} \left\{\int_T^{2T}dt \int_T^t \phi(x^{\prime},t^{\prime}) dt^{\prime} -
\int_0^{T}dt \int_0^t \phi(x^{\prime},t^{\prime}) dt^{\prime} +T\int_0^T \phi(x^{\prime},t) dt
\right\}\;.
\end{equation}

Applying twice the multiplicative integration rule, we obtain
\begin{equation}
\label{A7}
\mathcal{J}^E=\frac{1}{T^2}\int_0^L dx (L-x)
\int_0^T dt \left\{T\phi(x,t+T)-t\left[\phi(x,t+T)-\phi(x,t)\right]\right\}\;.
\end{equation}

By introducing the center of kink $\widetilde{X}(t)$ weighted by function $\phi(x,t)$ in analogy to~(\ref{A4}), and using
$\left[\widetilde{X}(t+T)-\widetilde{X}(t)\right]=\mathcal{V}T$, expression~(\ref{A7}) becomes
\begin{equation}
\label{A8}
\mathcal{J}^E=-\frac{1}{T}\int_0^T dt 
\int_0^L  \phi(x-\mathcal{V}T,t)\cdot(x-\mathcal{V}t) dx-
\frac{\mathcal{V}}{T}\int_0^T t dt 
\int_0^L \phi[\varphi^{v}(t)] dx
\;.
\end{equation}

Taking into account that $\langle \phi[\varphi^v(t)]\rangle_T\equiv 0$, we finally obtain expression~(\ref{JEXCH_red}) for the exchange current.

\end{document}